\documentclass[twocolumn,english]{revtex4}
\usepackage[T1]{fontenc}
\usepackage[latin9]{inputenc}
\usepackage{graphicx}
\usepackage{amssymb}

\makeatletter

\usepackage{babel}
\makeatother

\begin{document}

\title{Entanglement and the Speed of Evolution in Mixed States}

\author{Judy Kupferman and Benni Reznik}
\affiliation{ School of Physics and Astronomy, Raymond and Beverly Sackler Faculty of Exact Sciences, Tel Aviv University, Tel Aviv 69978, Israel. }
\begin{abstract}
Entanglement speeds up evolution of a pure bipartite spin state, in
line with the time energy uncertainty. However if the state is mixed
this is not necessarily the case. We provide a counter example and
point to other factors affecting evolution in mixed states, including
classical correlations and entropy.
\end{abstract}
\maketitle

\section{Introduction}

The speed with which a quantum system evolves from one state to the
next is a subject of both theoretical and practical interest. The
question sheds light on the nature of time and of quantum evolution
in general\cite{aharonov etc}. Practical applications are evident
in the design of a possible quantum computer, and also in quantum
metrology including the establishment of precise frequency standards
\cite{huelga}, and in optical and atomic clocks\cite{clocks}.
Some applications will require as fast a period as possible, while
others require maximal distinguishability between states. Since practical
applications nearly always involve some loss of purity, it is also
necessary to consider the evolution of mixed states.

Speed of evolution of a quantum system has been shown to be inversely
proportional to energy or energy spread ($E$ or $\Delta E$) \cite{margolis etc}.
It has also been shown that entanglement speeds up evolution\cite{giovanetti entanglement}.
We find that with mixed states these are not always the case. Mixed
states with the identical $\Delta E$ show completely different time
evolution, and there is no clear relationship to the entanglement.
In addition to entanglement, factors related to the mixing contribute.

We look at a specific model of a bipartite two level system system
and examine factors affecting time evolution, first when the system
is in a pure state, and then when it is mixed. With a pure state the
system behaves as expected. However with mixed states we find surprising
behavior. We look at three states which mix a maximally entangled
state and separable states in different representative ways. One is
a Werner state, which mixes a maximally entangled state with a maximally
mixed state: $\rho_{wer}=\frac{1-x}{4}\, Id\,+\, x\left|\Psi^{+}\right\rangle \left\langle \Psi^{+}\right|.$
where $\left|\Psi^{+}\right\rangle =\frac{1}{\sqrt{2}}\left(\left|10\right\rangle +\left|01\right\rangle \right)$.
The second, which we nickname $\rho_{gisin}$ \cite{gisin}, mixes
$\Psi^{+}$ with a mixture of product states: $\rho_{gis}=\frac{(1-x)}{2}\,\left[\left|00\right\rangle \left\langle 00\right|+\left|11\right\rangle \left\langle 11\right|\right]\,\,+\, x\left|\Psi^{+}\right\rangle \left\langle \Psi^{+}\right|.$
This will enable us to inspect the role of classical correlations
of the product states. The third state is a mixture of $\Psi^{+}$
with a maximally polarized state with no correlations at all: $\rho_{3}=\left(1-x\right)\left|00\right\rangle \left\langle 00\right|+x\left|\Psi^{+}\right\rangle \left\langle \Psi^{+}\right|.$
$\rho_{wer}$ is entangled for $x\geq\frac{1}{3}$, $\rho_{gis}$
for $x\geq\frac{1}{2}$, and $\rho_{3}$ for all $x>0$.

\begin{figure}[h]
\includegraphics[scale=0.25]{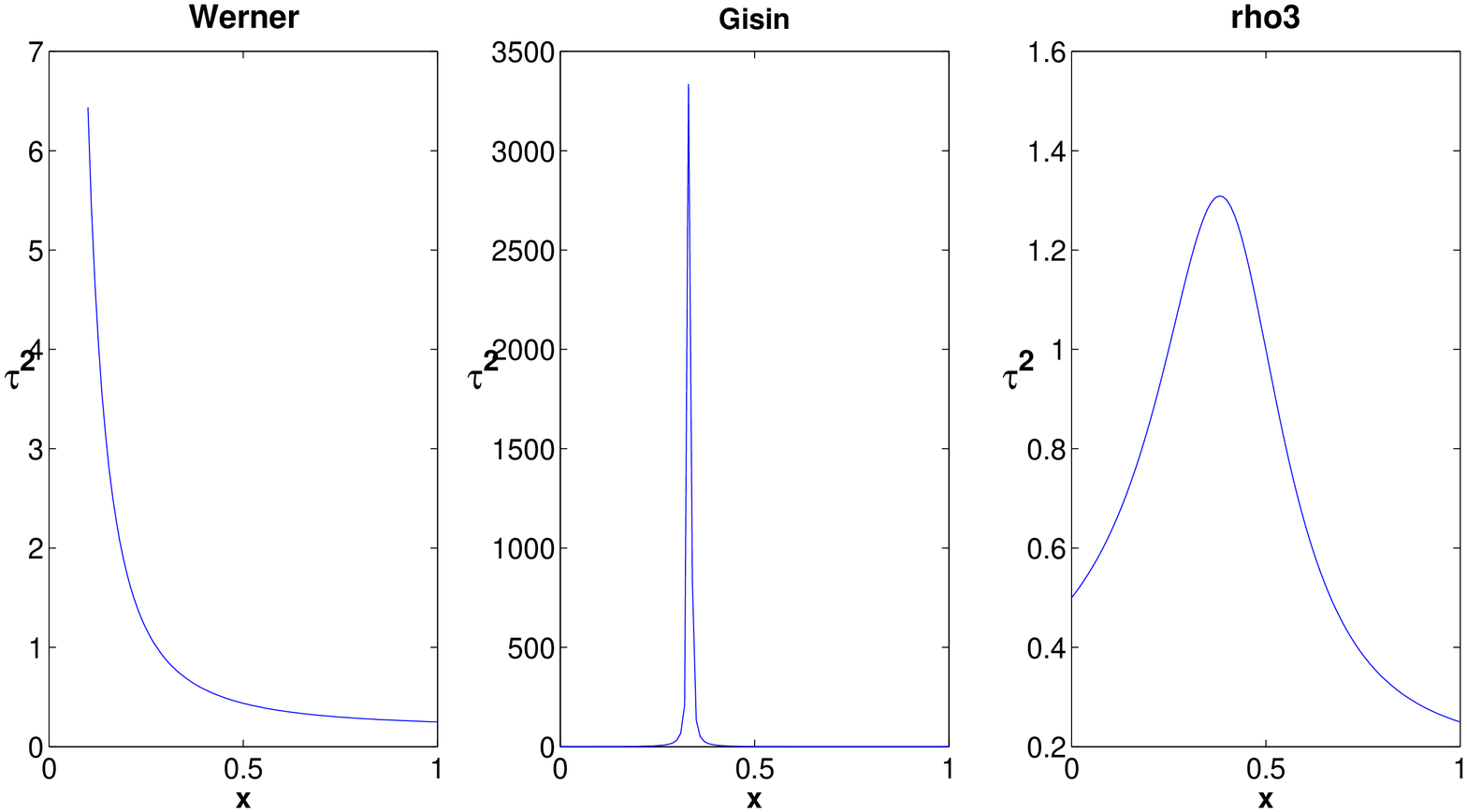}

\caption{\label{fig:Tausquared-for-the}Time evolution for the three matrices
(decay coefficient as function of mixing)}

\end{figure}

Our model has two physically separated spins each in a magnetic field,
with the field for each spin orthogonal to the spin axis, so that
the evolution operator contains $\sigma_{x}$ for each spin. Figure
\ref{fig:Tausquared-for-the} shows time evolution for these three
classes of mixed state for small t. The graphs show the speed of decay
from the original state as a function of x. For all three classes
of states, taking $\left\langle E\right\rangle =tr\left(\rho H\right)$
we find $\Delta E=2(1+x)$ . However we see in Figure \ref{fig:Tausquared-for-the}
that although \emph{the states have the} \emph{same} $\Delta E$ ,
their \emph{time evolution is different.} 

As will be discussed in Section \ref{sec:Mixed-states:-short}, the
change in a state is proportional to $exp\left(-t^{2}/\tau^{2}\right)$
, and Figure \ref{fig:Tausquared-for-the} shows $\tau^{2}$ as a
function of $x$, so that the lower the graph, the faster the rate
of decay. With the Werner state entanglement increases with x, and
indeed the state is stationary at the identity and speed of evolution
increases with increasing x. However the center graph of the Gisin
state shows increasing speed both for increasing x (increasing entanglement),
and also with \emph{decreasing} x. For the Gisin state, decreasing
x means increasing classical correlations so that these too may affect
the speed. The third matrix gives the most surprising graph. Entanglement,
as measured by the concurrence\cite{wooters}, rises linearly with
x, but the graph of the speed of decay is clearly not a monotonic
function of the entanglement and for $x<0.38$ speed increases as
entanglement decreases. Thus we see from Figure \ref{fig:Tausquared-for-the}
that short time evolution reveals the essential difference between
mixed and pure states: with mixed states speed is no longer inversely
proportional to $\Delta E$, and it is not necessarily a monotonic
function of entanglement. 

Therefore other factors are at work as well. These are connected to
the mixing and include classical correlations and entropy. In this
paper we provide an analysis of their contribution to evolution. The
paper is organized as follows. First we give details of the model
and its evolution when the system is in a pure state. Then we consider
mixed states. In that case it becomes necessary to inspect different
aspects of evolution separately: evolution for a short time (as discussed
above), period of evolution and maximal distance from the original
state. We then separate the effect of classical correlations by looking
at the period and distance of states which are mixed but not entangled.
We see that classical correlations too can speed up the period of
evolution, but at the expense of reducing maximal distance. This can
be useful in applications where the change itself is important rather
than the distance between states. Finally, we show through numerical
means that all correlations speed up evolution while entropy slows
it down.

\section{\label{sec:Details-of-the}Details of the model}

We take a bipartite two level system, modeled as two spins in magnetic
fields. The spins are physically separated so that a local operation
on one will not locally affect the other. The Hamiltonian for the
total system is $H=H_{a}+H_{b},$ where $H_{i}=\vec{\sigma_{i}\cdot}\hat{n_{i}}$
, $i$ refers to the first or second spin and $\hat{n_{i}}$ refers
to the direction of the magnetic field at the location of that spin.

For a pure state it is easily shown that speed of evolution is a monotonic
rising function of entanglement. Taking a state with arbitrary entanglement,
$\Psi=\alpha\left|\uparrow\uparrow\right\rangle +\beta\left|\downarrow\downarrow\right\rangle $,
where quantization is along the z axis and entanglement is maximal
for $\alpha=\beta=\frac{1}{\sqrt{2}}$, the state will reach orthogonality
when $\left\langle \Psi\right|U\left|\Psi\right\rangle =0,$ where
$U=e^{-i\left(\vec{\sigma}\cdot\hat{n}\right)_{a}+\left(\vec{\sigma}\cdot\hat{n}\right)_{b}}=\left(\cos t\cdot Id-i\sin t\vec{\sigma}\cdot\hat{n}\right)_{a}\left(\cos t\cdot Id-i\sin t\vec{\sigma}\cdot\hat{n}\right)_{b}$
. Equating the real and imaginary parts separately to zero we find
the time to reach an orthogonal state, \begin{equation}
t_{\perp}=arc\cot\sqrt{\left\langle \left(\vec{\sigma_{a}}\cdot\hat{n_{a}}\right)\cdot\left(\vec{\sigma}_{b}\cdot\hat{n}_{b}\right)\right\rangle }.\end{equation}
 Plugging in, for example, $\sigma_{x}$ on both spins, $t_{\perp}=\frac{\pi}{2}$
for a product state, where either $\alpha$ or $\beta$ is zero. The
time decreases continuously with increasing entanglement, and for
a maximally entangled state reaches a minimum at $\frac{\pi}{4}$.

The magnet angles which will give maximal speed are obtained by minimizing
the function for time, and by taking into account constraints from
the imaginary part of the equation. It turns out that optimal angles
for product states must have one of the magnets orthogonal to its
spin axis, while for maximally entangled states the optimum depends
solely on the relationship between the two magnets. For example for
the anticorrelated singlet optimal magnet angles are antilinear, whereas
for a correlated entangled state $\frac{1}{\sqrt{2}}\left(\left|11\right\rangle +\left|00\right\rangle \right)$
we have $\theta_{a}=\theta_{b},\,\phi=-\phi_{b}$. Each of the four
Bell states has a different set of constraints for optimal angles,
so that we have an external physical constraint reflecting the inner
structure of the state.

\section{Mixed states: short term evolution\label{sec:Mixed-states:-short}}

Time evolution in this model has (at least) three aspects: evolution
for a short time, period, and maximal distance from the original state.
With pure states this distinction does not add information, but in
the case of mixed states each of the three aspects gives different
results.

We begin with evolution for a short time, which we call the kickoff.
For pure states {\footnotesize \begin{eqnarray}
\left|\left\langle \Psi_{0}\mid\Psi(t)\right\rangle \right|^{2} & = & \left|\left\langle \Psi_{0}\mid e^{-\frac{i}{\hbar}Ht}\mid\Psi_{0}\right\rangle \right|^{2}\nonumber \\
 & \simeq & \left|1-\frac{it}{\hbar}\left\langle H\right\rangle -\frac{t^{2}}{2\hbar^{2}}\left\langle H^{2}\right\rangle \right|^{2}\nonumber \\
 & = & 1-\frac{t^{2}}{\hbar^{2}}\left(\left\langle H^{2}\right\rangle -\left\langle H\right\rangle ^{2}\right)\cong e^{-\frac{t^{2}}{\hbar^{2}}\delta E^{2}}\label{eq:frol}\end{eqnarray}
}so that for a short time the decay coefficient is $\delta E^{2}/\hbar^{2}$
and the speed of evolution is indeed proportional to the energy spread.

For mixed states we take a direct analog of fidelity in the density
matrix formalism: {\footnotesize \begin{equation}
F\left(\rho\left(t=0\right),\rho(t)\right)=\frac{tr\left(\rho\left(t=0\right)\cdot\rho(t)\right)}{tr\left(\rho\left(t=0\right)^{2}\right)}=\frac{tr\left(\rho_{0}U\rho_{0}U^{\dagger}\right)}{tr(\rho_{0})^{2}}\label{eq:dm fidelity}\end{equation}
}The denominator is for normalization. The transparent analogy to
fidelity for pure states makes this a convenient measure: for a pure
state where $\rho=\left|\Psi\right\rangle \left\langle \Psi\right|$
this reduces to{\footnotesize \begin{equation}
\frac{\left\langle \Psi\right|\left|\Psi\right\rangle \left\langle \Psi\right|U\left|\Psi\right\rangle \left\langle \Psi\right|U^{\dagger}\left|\Psi\right\rangle }{\left\langle \Psi\right|\left|\Psi\right\rangle }\end{equation}
}which is just the square of the fidelity. Expanding $F$ as we did
in Eq.\ref{eq:frol} by taking $U\simeq1-iHt$ we obtain{\footnotesize \begin{equation}
\frac{tr\left(\rho_{0}U\rho_{0}U^{\dagger}\right)}{tr(\rho_{0}^{2})}=1-t^{2}\frac{tr\rho\rho HH-tr\rho H\rho H}{tr\rho^{2}}\simeq e^{-\frac{t^{2}}{\tau^{2}}}\end{equation}
}so that (taking $\hbar=1),$ $\frac{1}{\tau^{2}}$ is analogous to
$\delta E^{2}$ for pure states. Therefore we can examine the relationship
between energy spread and time in the case of density matrices as
well. Figure \ref{fig:Tausquared-for-the} shows $\tau^{2}$ as a
function of $x$. Though $\delta E^{2}$ is identical for all three
classes of states: $\rho_{wer}$, $\rho_{gis}$ and $\rho_{3}$ ,
they have clearly different behavior \cite{note1}.

\begin{figure}[h]
\includegraphics[scale=0.22]{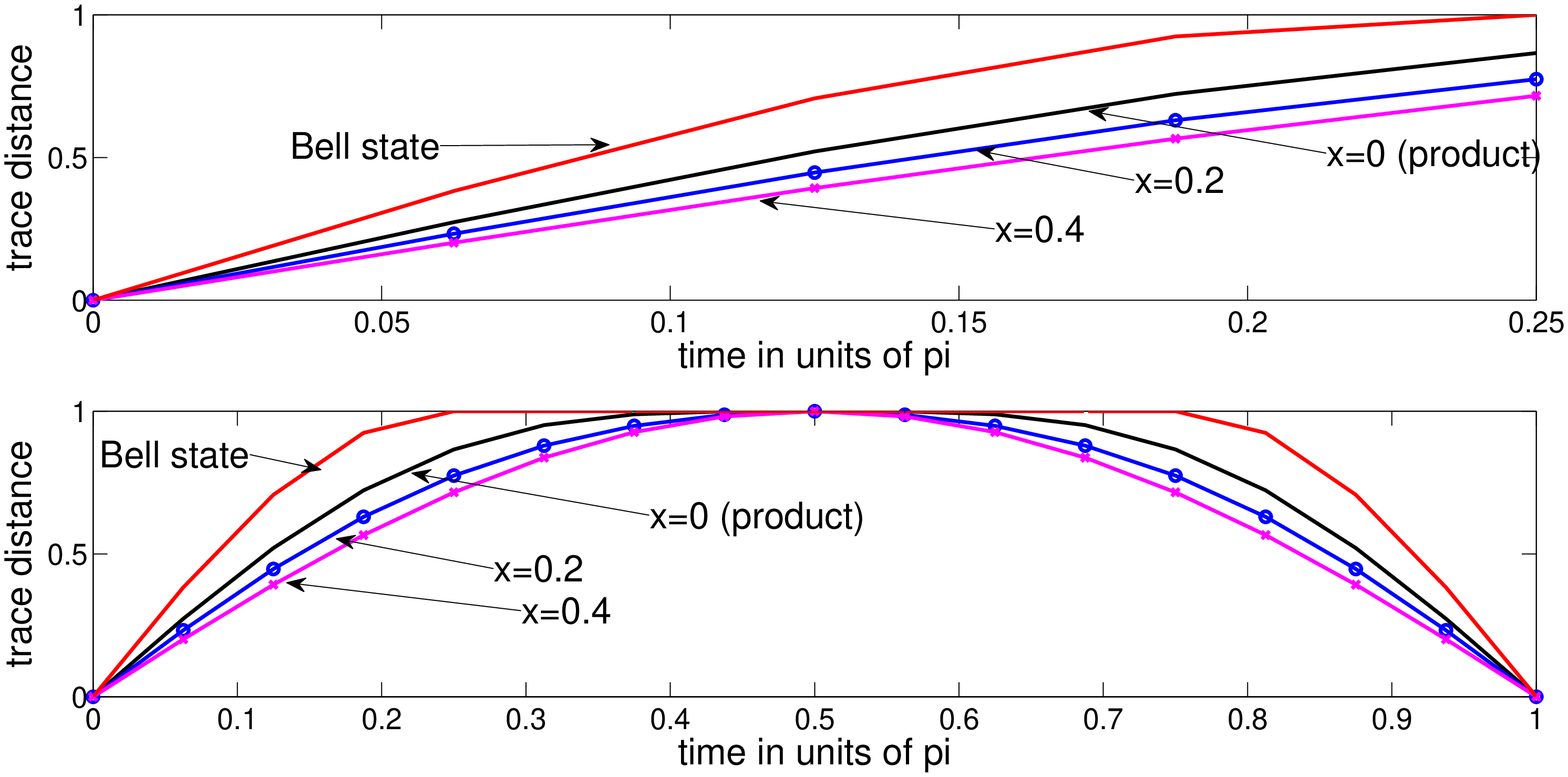}

\caption{\label{fig:trdist_partent_mixed}Trace distance for partially entangled
mixed state $\rho_{3}$. The concurrence grows linearly with x. Top
is for time till $\frac{\pi}{4}$ and bottom shows a full cycle. The
lowest curve in both graphs is more entangled than the second lowest.}

\end{figure}

This behavior is also seen with the more conventional trace distance
measure. Trace distance is defined as $D(\rho,\sigma)\equiv\frac{1}{2}tr\left|\rho-\sigma\right|$\cite{nielsen chuang}.
We take $\rho=\rho(t=0)$ and $\sigma=\rho(t)=U\rho U^{\dagger}.$
For pure states this gives the same result as in Section \ref{sec:Details-of-the}:
the more entangled the initial state, the faster it evolves to an
orthogonal state. The trace distance for a pure general entangled
state $\left|\psi_{ent}\right\rangle =\cos\gamma\left|10\right\rangle -\sin\gamma\left|01\right\rangle $
is $D\left(t\right)=\left|\sin\left(2\gamma\right)\sin\left(2t\right)\right|$
. So at $t=\frac{\pi}{4}$ the distance may equal unity (orthogonal
state), but only for maximal entanglement where $\sin\gamma=\frac{1}{\sqrt{2}}$
; for lower entanglement, orthogonality is reached at a later time.
For mixed states the graph of trace distance for small t shows the
same surprising results for our three matrices as that shown in Figure
\ref{fig:Tausquared-for-the}. For example with $\rho_{3}$, for a
certain range of values of x, states which are more entangled evolve
\emph{more slowly} than states which are less entangled. See Figure
\ref{fig:trdist_partent_mixed}.

As before, we see that contrary to expectation, there is no clear
indication that the speed increases as a result of increasing entanglement:
or rather, it increases with increasing entanglement but also in other
cases. Entanglement may be affecting the speed but there must be \emph{another}
factor or factors as well \cite{note2}. The question is what these factors are, aside from energy and energy
spread. The additional factors must be related to the mixing, and
include classical correlations and entropy.

\section{non entangled mixed states }

The other two aspects of time evolution are helpful in unveiling additional
factors, and it is important to distinguish between them. Time evolution
in this model is periodic. We therefore look at two different aspects
of period. First is the length of a half cycle, that is, the time
needed to reach an orthogonal or maximally distant state, which for
simplicity we will refer to as {}``period'' The second aspect of
interest is the maximal achievable distance. Both of these are seen
with the trace distance measure, and optimizing for each gives different
results.

The problem with evolution of mixed states is to separate out the
different effects of mixing and of classical and quantum correlations.
In order to isolate the effect of classical factors we first examine
a mixed state with no quantum correlations at all. In this section
we treat a specific toy model, in order to clarify the different effects
of period and distance. In the next section we will deal with separable
states in general, and attempt to isolate the factors of correlation
and entropy.

Quantum correlations were expressed in Section \ref{sec:Details-of-the}
using $\alpha$ and $\beta$, which went from $0$ (or $1$) for a
product state to $\frac{1}{\sqrt{2}}$ for maximal entanglement. We
now look at a classical analogy in a non-entangled density matrix,
$\rho=a\left|11\right\rangle \left\langle 11\right|+(1-a)\left|00\right\rangle \left\langle 00\right|$,
where for $a=\frac{1}{2}$ we call the state maximally correlated
in analogy to the quantum formalism. This is not a standard measure
of correlations; that would be better expressed with mutual information,
and we do so in Section \ref{sec:Mutual-information-and}. The measure
in this section attempts to look only at the extent of the departure
from a pure product state to a mixed correlated state, and its effect
on time evolution.

We look at the time evolution from two different aspects. First we
optimize for period, choosing magnet angles which give the fastest
possible period regardless of the distance achieved. Afterwards we
optimize for trace distance rather than period. In the first case
the mixing leads to speedup and reduces distance. In the second it
causes slowdown but eventually all states reach orthogonality.

\begin{figure}[h]
\emph{\includegraphics[scale=0.22]{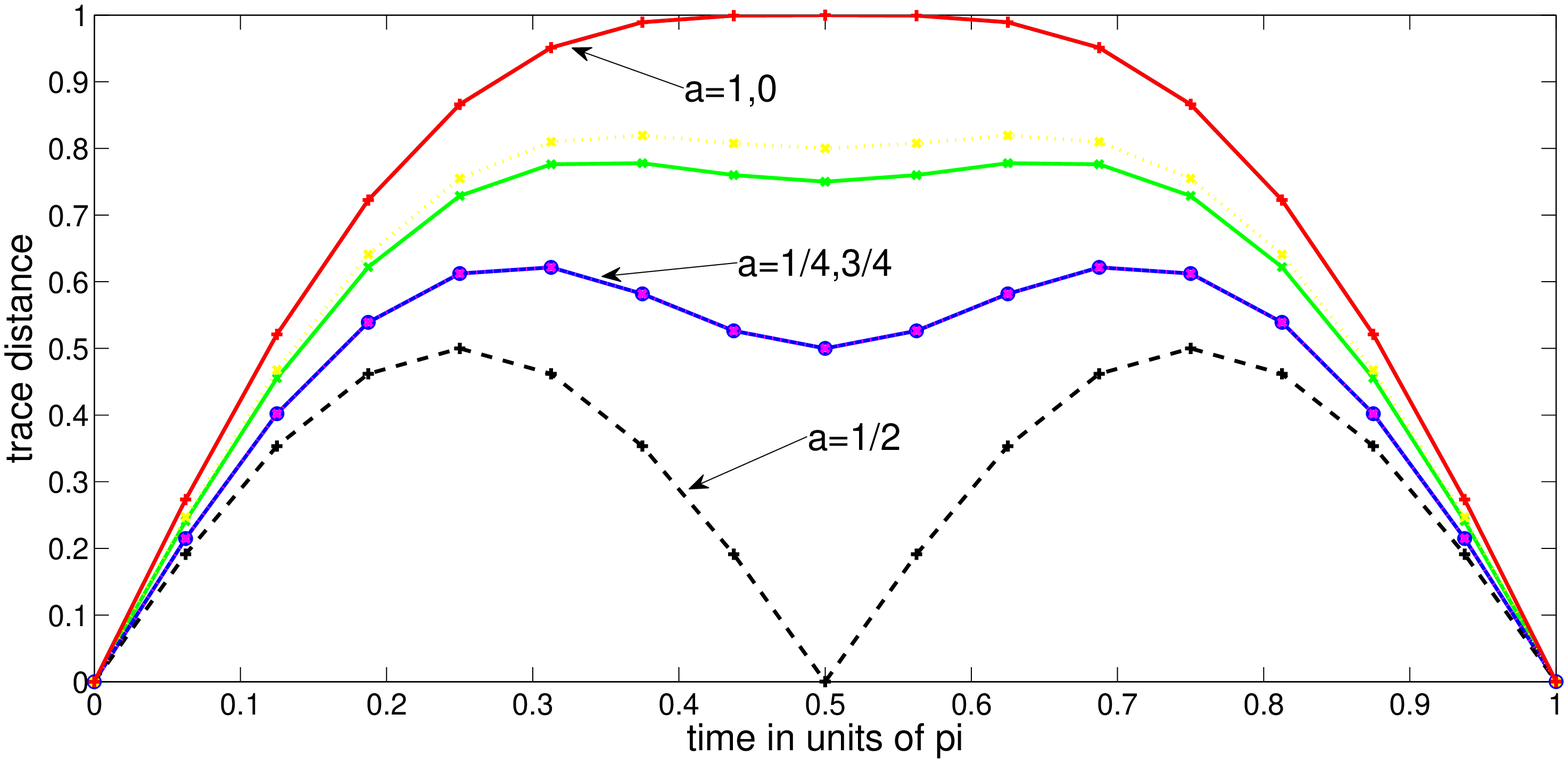}}\caption{\label{fig:prodmix dist for dif x's} (Color online) Trace distance
of product mixture for various values of a ($\sigma_{x}$ on both
spins). Top curve (red): a=0,1. Lowest curve (black): a=1/2. Second
lowest curve(blue, magenta): a=1/4, 3/4. Third lowest curve (green):
a=1/8. Second highest curve (yellow): a=9/10.}

\end{figure}

Figure \ref{fig:prodmix dist for dif x's} shows a graph of the trace
distance as a function of time for representative values of a, optimized
for period. %
 The fastest period ($\frac{\pi}{4}$ ) but with the smallest achievable
distance ($\frac{1}{2}$) are attained when the state is maximally
mixed, $a=\frac{1}{2}$ . For a pure product state, $a=1$ and $a=0$,
the period is twice as slow $\left(\frac{\pi}{2}\right)$ but the
states reach orthogonality with a distance of 1. The graph shows that
as the mixing approaches the product state, the point of maximal distance
comes later and the distance grows. Thus we see that under the appropriate
operators increasing mixing reduces trace distance, but the distance
reaches maximum more quickly.

\begin{figure}[h]
\includegraphics[scale=0.22]{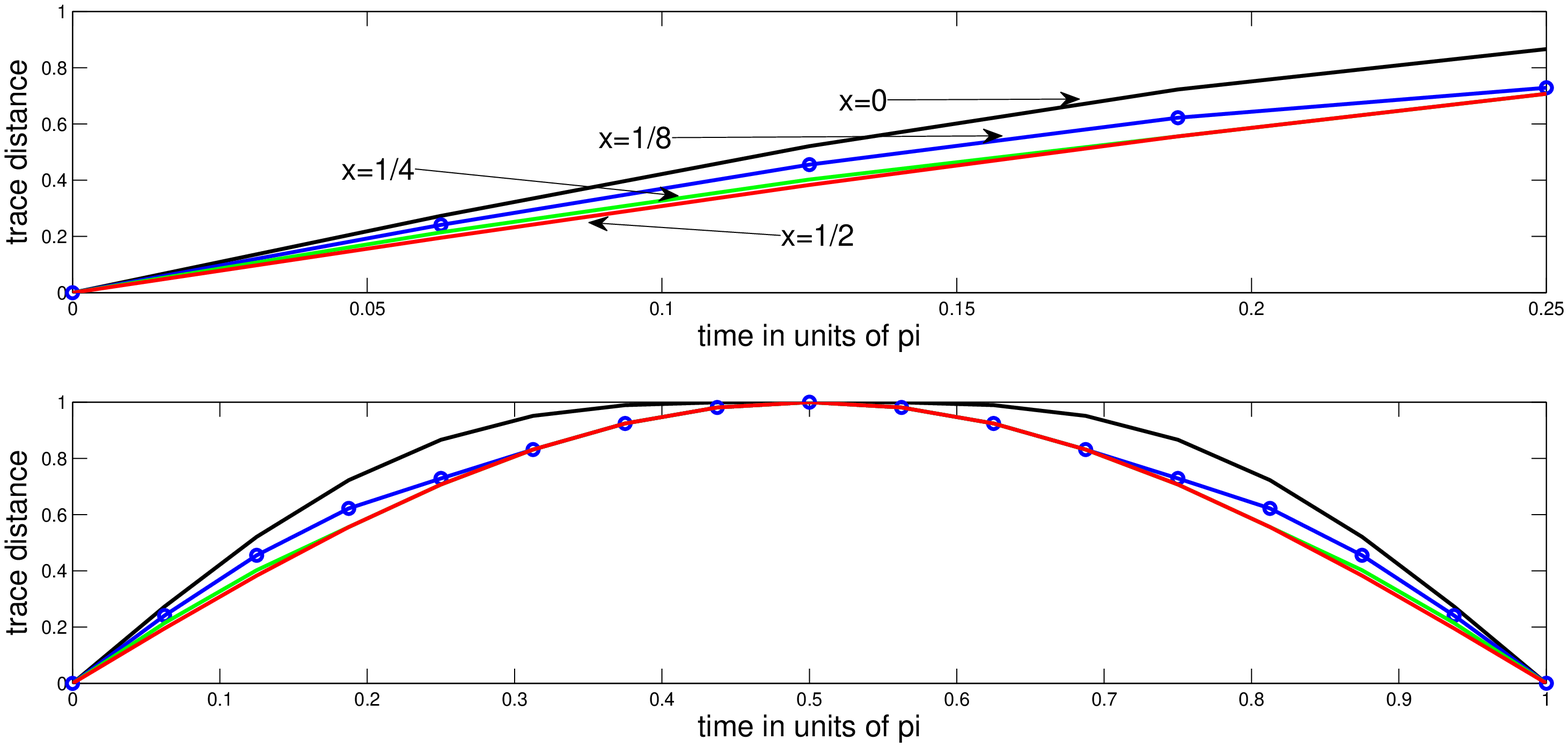}

\caption{\label{fig:trdist_prodmix}(Color online) Trace distance for product
mixture, optimized for maximum distance. Top is for time till $\frac{\pi}{4}$
and bottom shows a full cycle. In both graphs the highest (black)
curve is a pure product state, the lowest is maximally mixed ($a=\frac{1}{2}$),
and slope increases as a approaches 0 or 1. Note, however, that all
cases reach an orthogonal state at the same time $\frac{\pi}{2}$
.}

\end{figure}

If instead, we optimize angles for the largest attainable trace distance
(which is at the expense of speed) we see in Figure \ref{fig:trdist_prodmix}
that all states achieve optimal distance, but the greater the mixing
(the farther from a pure product state), the slower the kickoff and
general rate of evolution. %
 This is probably the reason that when optimized for trace distance,
entangled mixed states do not necessarily achieve the optimal distance
so quickly: the mixing attentuates the result. For practical purposes
it is the first case, with optimization for period, which is interesting:
in applications where the speed of change is important rather than
its extent, increasing mixing of classically correlated states may
improve the result. It should be noted that this applies only to a
bipartite state. For larger systems further investigation is necessary.

\section{\label{sec:Mutual-information-and}Mutual information and entropy}

\begin{figure}[h]
\includegraphics[scale=0.25]{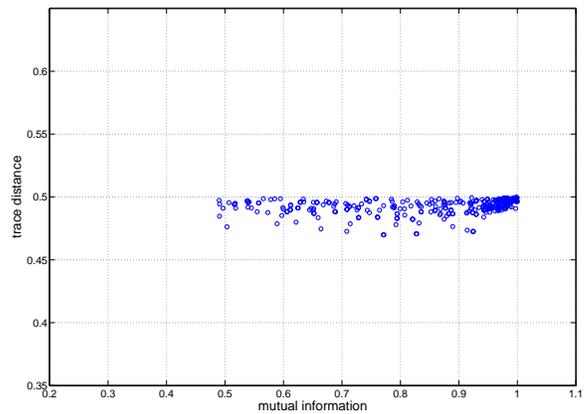}\caption{\label{fig:dist(mutualinfo)}Trace distance as a function of mutual
information for separable states. Most, but not all states which achieve
maximal distance have mutual information approaching 1. Mean: 0.8608(x),0.4928(y).
Median: 0.9296(x), 0.4942 (y). Std: 0.1461(x), 0.005979(y).}

\end{figure}

In the previous section we looked at the effect of mixing as a whole
on a toy model, and distinguished between the aspects of period and
distance. We now deal with the problem on a more general basis, and
attempt to separate out the factors of correlations and of entropy.
A measure of the total correlations between two subsystems is mutual
information, defined in terms of entropy: $I\left(\rho_{AB}\right)=S\left(\rho_{A}\right)+S\left(\rho_{B}\right)-S\left(\rho_{AB}\right)$.
This is actually the relative entropy between $\rho_{AB}$ and $\rho_{A}\otimes\rho_{B}$
and so it is a measure of correlations between the subsystems \cite{vedral}
. Groisman et al.\cite{berry} have shown that mutual information
may be seen as describing both classical and quantum correlations.
States with maximal classical correlations have I=1, and entangled
states include classical as well as quantum correlations and have
higher I, up to 2 for maximally entangled states. For separable states,
a numerical search went through a total of $36\times10^{6}$ separable
states. Nearly all of these separable states are mixed (pure states
constitute only the surface of the Bloch sphere.) For these states,
we show in Figure \ref{fig:dist(mutualinfo)} the difference in distance,
$D_{dif}=D\left(t=\frac{\pi}{4}\right)-D\left(t=\frac{\pi}{2}\right)$,
as a function of mutual information, where for each state we optimized
angles to give the highest value of $D_{dif}$. This optimizes for
fastest period, that is, reaching the maximal trace distance as quickly
as possible. Taking the distance from $0-\frac{\pi}{4}$ would be
deceptive, because we have seen that one state may have a higher distance
at $\frac{\pi}{4}$ than another, but then continue to a maximum at,
for example, $\frac{\pi}{3}$, so that it actually reaches its maximum
later, while the first state reached its maximum at $\frac{\pi}{4}$.
We could have optimized for kickoff or maximal distance rather than
period, of course; it was necessary to choose one specific aspect
of evolution, and this proved the least ambiguous of the three. In
the previous section when optimizing for period we found a specific
reference point that proves useful: a distance of $1/2$ for $t=\frac{\pi}{4}$.
Indeed we will see with a numerical search that no unentangled state
reaches a greater distance at this point.

We found that the fastest evolving states have a half-cycle of $\frac{\pi}{4}$
and so this function $D_{dif}$ shows the maximal distance obtained
at the fastest cycle. As with the toy model, here too the maximal
trace distance for this period is found to be 1/2. Therefore if correlations
are the sole cause we would expect all optimal states which have a
distance of 1/2 to have mutual information of 1, that is, maximal
total correlations for a separable state. In fact the majority of
these states do - but \emph{not all} of them. Some states which do
not have maximal correlations still achieve the optimum trace distance.
Therefore trace distance is not a monotonic function of the total
correlations, and some other factor is in effect as well.

\begin{figure}[h]
\includegraphics[scale=0.25]{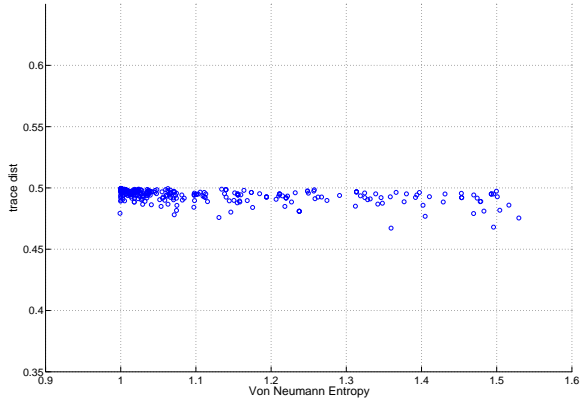}

\caption{\label{fig:distance_entropy}Trace distance as a function of entropy
for separable states. Most, but not all states which achieve maximal
distance have entropy approaching 1. Mean: 1.13(x),0.481(y). Median:
1.057(x), 0.4951 (y). Std: 0.1572(x), 0.09249(y).}

\end{figure}

We therefore used the results of the same search to graph trace distance
as a function of Von Neumann entropy. The result is similar to the
previous section: most states with the optimal trace distance have
the minimal entropy $S=1$, as for a maximally correlated state -
but again, not all of them. Some states with higher entropy also achieve
the maximum trace distance, as shown in Figure \ref{fig:distance_entropy}.
Yet although the same data were used for both - the same states each
with its trace distance - the graphs are not exactly inverse images
of each other. It appears that in nonentangled mixed states both correlations
and entropy affect time evolution and further work is necessary to
clarify the effect.

\section{Entangled mixed states}

We now turn to entangled states. It would be desirable to isolate
the effect of entanglement on evolution from the other factors in
all mixed states. At present we can show this for the three matrices
mentioned above, as follows:

\begin{figure}[h]
\includegraphics[scale=0.25]{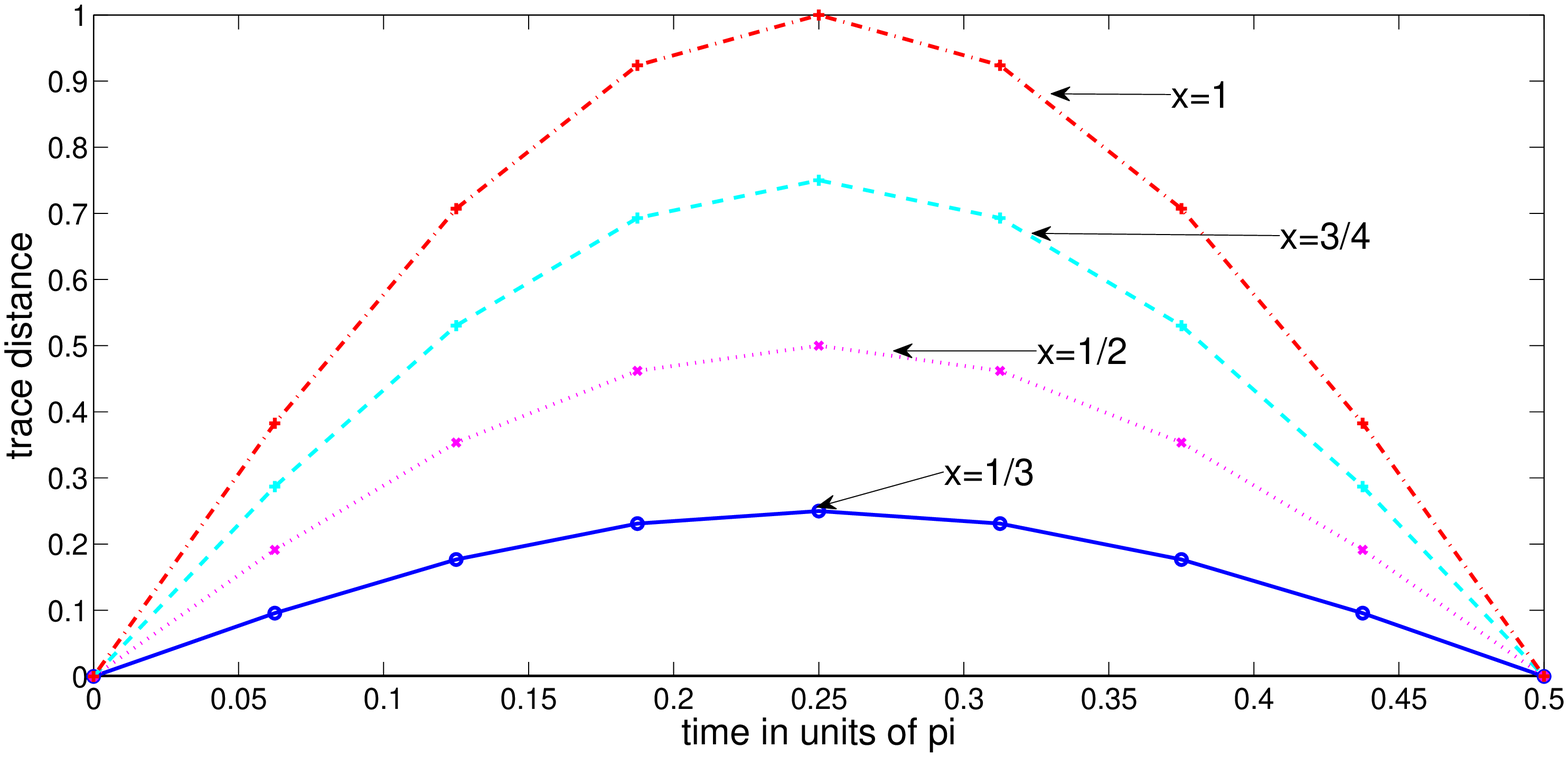}\caption{{\footnotesize \label{fig:(Color-online)D(t)for 3 peres states}(Color
online)Trace distance as function of time for the three representative
matrices. Magnet angles are $\sigma_{z},\sigma_{-z}$ . With these
angles, the graph is the same for all three matrices: $\rho_{wer}$,
$\rho_{Gisin}$ and $\rho_{3}$. The curves represent mixing values.
Lowest (blue) x=1/3, entangled for the Werner state, but not for $\rho_{Gisin}$;
highest (red): x=1, maximally entangled for all three matrices and
reaches a distance of 1, that is, an orthogonal state. For a product
mixture, $x\left|11\right\rangle \left\langle 11\right|+\left(1-x\right)\left|00\right\rangle \left\langle 00\right|$,
the trace distance is zero for all values of x.}}

\end{figure}

When magnet angles are $\hat{z}$, $-\hat{z}$ , as shown in Figure
\ref{fig:(Color-online)D(t)for 3 peres states}, all three matrices
have the same graph, while the product mixture does not evolve at
all. This is because product states (e.g.$\left|11\right\rangle \left\langle 11\right|$)
are invariant at these angles, so that in this column we are looking
\emph{only} at the effect of the amount of entanglement on the trace
distance and have \emph{neutralized the effect of mixing product states}.
In this case all three matrices give exactly the same analytical function
for trace distance, $D=|xsin(2t)|$ and thus double cycle is preserved
for all x , that is, for any degree of entanglement in the mixture.
In addition, x attenuates the maximum distance. It is notable that
three matrices have the identical analytical trace distance function,
even though the third is always entangled and the others are not.
In addition, the general shape is the same whether the matrix is entangled
or not (e.g. $x<\frac{1}{3}$ and $x>\frac{1}{3}$ for Werner states).
Entanglement clearly speeds up evolution, but the increase is smoothly
affected by something else as well and evidently this is not the mixing
with product states which has now been neutralized. It must be noted
that entangled states possess classical correlations as well. Therefore
evolution in this case may be affected both \emph{by quantum and by
classical} correlations. This would account for the smoothness of
the graphs.

However in addition all three matrices have the same energy spread
$\Delta E$ so the relation between energy spread and evolution seems
here to be in force. This shows that it is the mixing with product
states that distorted the relationship, and when they are neutralized
it returns.

\section{Conclusions}

We found that the speed of evolution in mixed states differs from
that in pure states in that it is influenced not only by entanglement
(which reflects the amount of energy spread) and by external constraints
- magnet angle in this model - but also by factors due to the mixing.
For the three matrices, the accepted $\Delta E\Delta t$ relationship
is preserved only if mixing with product states is neutralized. Mixing
introduces classical correlations and entropy, both of which affect
evolution. Neither of them alone can account for it, but the relationship
between them is not yet clear. This may be because the measure of
mutual information includes quantum correlations as well as classical,
and the two do not necessarily have the same effect (e.g. maximally
mixed classical correlations decrease distance with increased period,
quantum correlations do not). Further work might include a measure
which excludes quantum correlations and then a functional relationship
might be found between the effect of classical correlations and of
entropy. 

Period can be speeded up by maximally mixed classical correlation,
but this reduces trace distance. When optimized for maximal distance
rather than period, such correlations attenuate possible trace distance
and slow the evolution down. In any case it is necessary to clarify
which aspect of time evolution - period or maximal distance - is relevant
to the discussion, as optimization for either gives different results. 

In sum, time evolution in mixed states appears to be affected by entanglement,
classical correlations, entropy, and external constraints. We have
attempted to point out the various effects of these factors. These
conclusions are a result of numerical methods and of analytic calculations
for specific cases. Future work should include an attempt to reach
a general analytic expression for the relative contributions of classical
and quantum correlations as well as entropy to the speed of evolution. 

We would like to thank S. Marcovitch and F. Zapolsky for helpful discussions,
and S. Machnes whose QLib package for Matlab\cite{qlib} was very
useful for the numerical work. This research has been supported by
the Israel Science Foundation grant number 784/06.


\begin{thebibliography}{10}
\bibitem{note1}We also graphed $\tau^{2}$ for the three matrices with angles optimized
at each point of x; the graphs are less dramatic but display the same
general shape and slopes.
\end{thebibliography}
\end{document}